\documentclass[apj]{emulateapj}
\usepackage{amsmath}
\usepackage{natbib}

\begin{document}

\title{Broadband Photometry of $105$ Giant Arcs: Redshift 
Constraints and Implications for Giant Arc Statistics\altaffilmark{*}}

\author{Matthew B. Bayliss}

~~~~~

\affil{Department of Astronomy \& Astrophysics} \affil{Kavli Institute for
Cosmological Physics} \affil{The University of Chicago, Chicago, IL 60637}

~~~~

\affil{AND}

~~~~

\affil{Harvard-Smithsonian Center for Astrophysics} \affil{Department of 
Physics} \affil{Harvard University, Cambridge, MA 02138}

\email{mbayliss@cfa.harvard.edu}

\altaffiltext{*}{Based on observations taken at the Southern 
Astrophysical Research Telescope (SOAR), a collaboration between 
CNP-Brazil, NOAO, The University of North Carolina at Chapel Hill, 
and Michigan State University and the Canada-France-Hawaii 
Telescope (CFHT) on Mauna Kea, which is operated by the National 
Research Council of Canada, the Institut National des Sciences de 
l'Univers of the Centre National de la Recherche Scientifique of 
France, and the University of Hawaii.  Additional supporting observations 
come from the 2.5m Nordic Optical Telescope, operated on the island 
of La Palma jointly by Denmark, Finland, Iceland, Norway, and Sweden, 
in the Spanish Observatorio del Roque de los Muchachos of the Instituto de
Astrofisica de Canarias, and the Gemini Observatory, which is operated by the 
Association of Universities for Research in Astronomy, Inc., under a 
cooperative agreement with the NSF on behalf of the Gemini partnership: 
The United States, The United Kingdom, Canada, Chile, Australia, Brazil 
and Argentina}

\begin{abstract}

We measure the photometric properties of $105$ giant arcs that were identified in systematic searches for galaxy-cluster-scale strong lenses in the Second Red-Sequence Cluster Survey (RCS-2) and the Sloan Digital Sky Survey (SDSS). The cluster lenses span $0.2 < $~z$_{l} < 1.2$ in redshift, with a median $\bar{z}_{l}=0.58$. Using broadband color criteria we sort the entire arc sample into redshift bins based on $u-g$ and $g-r$ colors, and also $r-z$ colors for the $\sim90\%$ of arcs that have $z-$band data. This analysis yields broad redshift constraints with $71^{+5}_{-4}~\%$ of the arcs at z $\geq1.0$, $64^{+6}_{-4}~\%$ at z $\geq1.4$, $56^{+5}_{-4}~\%$  at z $\geq1.9$, and $21^{+4}_{-2}~\%$ at z $\geq2.7$. The remaining $29^{+03}_{-5}$~\% have z $<1$. The inferred median redshift is $\bar{z}_{s} = 2.0\pm0.1$, in good agreement with a previous determination from a smaller sample of brighter arcs ($g$~$\lesssim 22.5$). This agreement confirms that z$_{s}=2.0\pm0.1$ is the typical redshift for giant arcs with $g$~$\lesssim 24$ that are produced by cluster-scale strong lenses, and that there is no evidence for strong evolution in the redshift distribution of arcs over a wide range of $g-$band magnitudes ($20 \leq$~$g$~$\leq 24$). Establishing that half of all giant arcs are at z $\gtrsim2$ contributes significantly toward relieving the tension between the number of arcs observed and the number expected in a $\Lambda$CDM cosmology, but there is considerable evidence to  suggest that a discrepancy persists. Additionally, this work confirms that forthcoming large samples of giant arcs will supply the observational community with many magnified galaxies at z $\gtrsim2$.

\end{abstract}


\section{Introduction}
\label{sec:intro}

The statistics of strong gravitational lensing by clusters of galaxies 
provides a test of cosmological models, by comparing the observed abundance 
of giant arcs against predictions from ray-tracing cosmological N-body simulations. 
Incidents of strong lensing are often identifiable by the formation of giant arcs, 
which are background sources that 
are strongly lensed into multiple, often merging, images by a foreground 
gravitational potential. The frequency of giant arcs is a complex observable that scales
with the halo mass function, the detailed properties of high-redshift galaxy population, 
and the internal properties of the halos 
\citep[e.g.,][]{grossman1988,bartelmann1998,cooray1999,gilmore2009}. \citet{bartelmann1998} 
first suggested that the number of giants arcs on the sky is under-predicted by 
$\Lambda$CDM by approximately an order of magnitude, and later observational comparisons 
using small samples of giant arcs corroborated the apparent over-abundance of giant arcs 
on the sky compared to $\Lambda$CDM predictions 
\citep{luppino1999,zaritsky2003,gladders2003,li2006}. 
In principle, it should be possible to produce predictions for giant arc statistics that 
match what we observe, assuming that we possess a suitably accurate model describing 
the formation and evolution of structure over cosmic time. In practice, the apparent 
failure of predictions for giant arc statistics to match the observations by as much 
as an order of magnitude is clear evidence that more work remains to be done.

Resolving the persistent discrepancy between the abundance 
of giant arcs observed and the number predicted by early simulation efforts requires 
progress on two fronts: 1) improved empirical constraints from much 
larger samples of giant arcs that are identified in uniform data with a 
well-characterized selection, and 2) 
refined predictions based on simulated giant arc samples. It is tempting 
to consider the possibility of giant arc statistics providing direct observational 
evidence for divergence from the concordance $\Lambda$CDM cosmology 
\citep{fedeli2008,daloisio2011}, but we must first explore other physical explanations 
for the discrepancy between giant arc counts in observations and 
simulations. Motivated by the \citet{bartelmann1998} 
result, theoretical work over the past decade has explored a variety of 
effects which, if unaccounted for in theoretical predictions, can result in an 
under-abundance of simulated arcs compared to the number of giant arcs observed 
in the real universe. 

Suggested effects include the contribution of baryons to 
cluster lensing cross-sections in the form of central massive galaxies and substructure 
\citep{flores2000,menegh2000,menegh2003,hennawi2007,menegh2010}, the effects 
of dark matter being dragged into the cores of clusters by cooling baryons 
\citep{puchwein2005,rozo2008,wambs2008}, triaxiality of cluster mass profiles 
\citep{oguri2003,dalal2004,hennawi2007,menegh2010}, accounting for additional 
but uncorrelated structure in the universe that is projected along the line-of-sight 
of lens-source systems \citep{wambs2005,hilbert2007,puchwein2009}, short time-scale 
increases in the strong lensing cross-section due to major mergers 
\citep{torri2004,fedeli2006,hennawi2007}, and the properties of the assumed 
background galaxy source population \citep{hamana1997,wambs2004}. The 
aforementioned papers have generally quantified the impact of various effects in the 
context of comparing predictions and observations for the frequency of the formation 
of giant arcs, with several authors making arguments to disclaim the discrepancy 
noted by \citet{bartelmann1998} \citep[e.g.,][]{wambs2004,horesh2005,horesh2011}. 
These arguments, however, are based on the identification of {\it possible} mechanisms 
for explaining or alleviating the apparent giant arc discrepancy, and there is a persistent 
lack of empirical evidence to confirm or deny the relevance of these different mechanisms.

\subsection{The Background Source Population}

In this paper we focus on the background galaxy population as an input into predictions 
for giant arc abundances in simulations, and make a direct measurements of the giant arc 
redshift distribution for a large sample of real giant arcs. Evidence from theoretical 
work \citep{wambs2004} suggests that the use of appropriate distribution of background 
galaxies may have the greatest potential to explain the dramatic discrepancy between 
the giant arc counts observed and those predicted by simulations in a $\Lambda$CDM 
cosmology. The redshift distribution of background sources impacts the global efficiency 
for giant arc production through the angular diameter distance term in the gravitational 
lens equation \citep[Equ. 13,][]{narayan1996}, which has an explicit dependence on 
the ratio of the angular diameter distances between the lens and the source, and the observer 
and the source. In practice, this means that the critical surface mass density for a given 
foreground lens is a function of the source redshift, where higher source redshifts 
result in lower critical surface mass densities. When analyzing the total cross section 
for strong lensing of all clusters within some cosmological volume, a background source 
population that is shifted to higher redshifts will therefore require lower surface mass 
densities of foreground structures in order to become supercritical (i.e. to be strong 
lenses). There has, however, been little effort made to systematically measure the background 
source redshift distribution of a well-defined sample of arcs until very recently. 
\citet{bayliss2011a} first measured the redshift distribution for a spectroscopically complete 
sample of $28$ giant arcs that were observed with Gemini-South, and they demonstrate that the 
giant arc redshift distribution can only be reliably measured using a large sample of uniformly 
selected giant arcs. 

\citet{wambs2004} note an increase in the total strong lensing cross-section for 
galaxy cluster-scale lenses of a factor of $\sim10-20$ when the background sources that 
are lensed into giant arcs are assumed to be at corresponding source planes of z$_{s}=1.5-2$ 
rather than z$_{s}=1$. Other work in which this effect is quantified finds smaller increases 
in the strong lensing cross-section by a factor of $\sim3$ or so 
\citep{dalal2004,li2005,fedeli2006}, but this still implies a significant potential boost 
in the total cross-section for giant arc production if giant arcs are typically formed 
by lensing of galaxies at z$_{s}=2$, rather than the z$_{s}=1$ source population used by 
\citet{bartelmann1998}. The importance of understanding the properties of the 
background source population that is lensed into arcs is further highlighted by more 
recent simulation work to generate predictions for giant arc statistics.

In agreement with early semi-analytic arguments 
\citep{oguri2003}, \citet{hennawi2007} identify the dominant uncertainty in their 
ability to robustly predict giant arc counts to be the poor constraints on the 
number density of galaxies at high redshift that are available to be strongly lensed 
by foreground cluster lenses. This uncertainty is fundamentally a reflection of our ignorance 
of the faint tail of the surface brightness function of galaxies at high redshifts. 
When we consider the implications of \citet{wambs2004} and \citet{hennawi2007} together it 
becomes clear that we cannot realistically make any useful statements about cosmology 
from tests of giant arc statistics without a firm understanding of the population of 
galaxies that are typically lensed into giant arcs. 

The first spectroscopic measurement of the redshift distribution for a complete sample 
of well-selected giant arcs indicates a median arc redshift of $\bar{z}_{s} = 1.82$ 
\citep{bayliss2011a}. However, the sample used in that work was still relatively 
small (N$_{arcs}=28$), and potentially biased toward high-redshift 
giant arcs due to a target selection criteria which favored larger giant arc radii 
(R$_{arc}$) systems, where R$_{arc}$ is the average angular distance between an arc 
and the centroid of the foreground lensing potential. There should be a net 
correlation between R$_{arc}$ and the intrinsic Einstein radius, R$_{E}$, of a given lens, 
and for a given lens the Einstein radius grows monotonically with the redshift of the lensed 
source.  However, there should also be a large scatter in the correlation between R$_{arc}$ and 
R$_{E}$ that could easily be washed out by other factors such as the shape of the critical 
curves for each individual lens. It is not clear to what degree this potential selection bias 
should impact the results of \citet{bayliss2011a}, 
but there is a somewhat reasonable argument to be made that the redshift distribution 
measurement in \citet{bayliss2011a} could be biased significantly high. In order to ensure a 
robust understanding of the properties of the galaxies that are typically lensed into giant 
arcs, it is important that we extend these investigations further to incorporate very large 
sample of arcs that 
are not subject to the same selection effects as those in \citet{bayliss2011a}. However, 
measuring spectroscopic redshifts for a sample of $\sim100$ giant arcs is extremely 
observationally expensive, especially in a scenario where spectroscopic completeness (i.e. 
redshift measurements for $\sim100\%$ of the observed sample) is important.

We also note that counting giant arcs may be in large part free from the issue of magnification 
bias that is known to affect, for example, counts of unresolved galaxies or quasars behind 
foreground lensing structures \citep{broadhurst1995}. As \citet{dalal2004} point out, the 
magnification due to gravitational lensing is spatial in nature, with surface brightness a conserved 
quantity. The magnification does not, therefore, impact the number density of spatially resolved 
objects on the sky.  Giant arcs are always, by definition, spatially resolved along at least one 
(and sometimes both) axis(es) on the sky. Assuming that magnification bias does not strongly 
influence counts of giant arcs, we make no attempt to model or correct for the affect in this 
paper. In support of this assumption we note that $\S$~\ref{sec:context} includes a direct 
comparison of the median redshifts of two samples of giant arcs that were selected from imaging 
data that different in depth by $\sim1.5-2$ magnitudes. The observed similarity in the median 
redshifts of these two samples provides some empirical evidence to support the argument that 
magnification bias does not strongly impact surveys of giant arcs, but future arc searches that 
probe to fainter magnitudes will be necessary to robustly quantify the precise effect of the 
magnification bias on giant arc counts.

\subsection{Measuring the Properties of A New Giant Arc Sample}

The ``giant arc problem'' posed by the apparent excess of giant arcs on the 
sky relative to $\Lambda$CDM predictions persists because of limitations on both 
the theoretical predictions and observational constraints. Making serious progress 
toward addressing the issue requires improving the fidelity of the 
machinery for theoretical predictions, which has been progressing steadily in recent 
years, but it also demands improvement on the observational side with respect to defining 
useful observational samples for comparison against the best predictions that theory 
has to offer. To that end, we have undertaken a systematic search for giant arcs around optically 
selected clusters in two large imaging surveys: the Second Red Sequence Cluster Survey 
\citep[RCS-2;][]{gilbank2011} and the Sloan Digital Sky Survey \citep{york2000}. 
An exhaustive visual inspection of these two surveys has produced a combined sample 
of hundreds of uniformly selected giant arcs (M. D. Gladders et al. 
2011, in preparation; M. B. Bayliss et al. 2011, in preparation). Complete 
definition of these samples (completeness, purity, effective area/volume probed) 
will be forthcoming in future work, but even prior to the completion of the sample 
definition there is a tremendous amount to be learned by characterizing the properties 
of the giant arcs themselves. 

In this paper we use photometric color criteria to sort a uniformly-selected sample of 
$105$ arcs into four redshifts bins at z $>1$, and evaluate the implications of the 
resulting arc redshift distribution. The redshifts and linear arc brightnesses of the 
sample are two observables that can be measured directly from the data, and used to 
inform future efforts to generate simulation-based predictions of giant arc abundances 
and match them to our new large giant arc samples. As discussed above, we can also provide 
direct insight regarding a possible resolution of the ``giant arc problem'' by definitively 
establishing that most giant arcs are galaxies that reside at high redshift 
(i.e. z $\gtrsim2$). 

This paper is organized as follows: in $\S$~\ref{sec:obs} we summarize the photometric and 
spectroscopic data analyzed in this paper. $\S$~\ref{sec:methods} describes our analysis methods, 
including aperture photometry of giant arcs and the color-based sorting of giant arcs 
into broad redshift bins. In $\S$~\ref{sec:discussion} we summarize the state of the current 
literature regarding investigations into possible contributing factors to the ``giant arc 
problem'', and discuss the implications of our results.

All magnitudes presented in this paper are calibrated relative to the AB system, via 
the SDSS.

\begin{figure}
\centering
\includegraphics[scale=0.55]{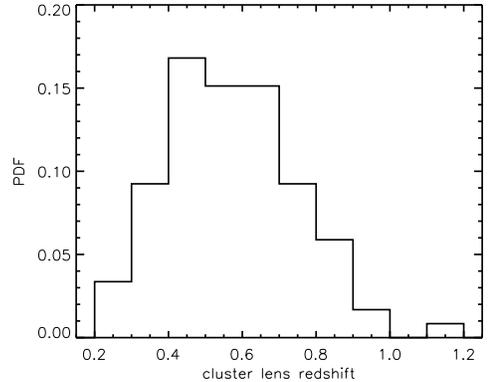}
\caption{\scriptsize{The probability distribution of photometric redshifts for the
population of foreground galaxy clusters that are responsible for producing
the giant arcs analyzed in this paper.}}
\label{lenses}
\end{figure}

\section{Observations}
\label{sec:obs}

\subsection{Giant Arcs Targeted for $u-$band Observations}
\label{sec:sample}

The objects analyzed in this paper are drawn from a pair of comprehensive 
surveys for giant arcs in the RCS-2 and SDSS survey imaging data. As a part of a 
larger collaboration, we have systematically searched the RCS-2 and SDSS DR7 
\citep{sdssdr7} for incidences of strong 
lensing by galaxy clusters. The search is performed by identifying 
lines-of-sight in the optical survey data which are likely to contain 
galaxy clusters using the red-sequence algorithm \citep{gladders2000,gladders2005}. 
The red-sequence cluster finding identifies over-densities of red galaxies on the 
sky, and provides photometric redshift estimates for the galaxy clusters from the 
redshifted $4000$\AA~ break. The galaxy cluster samples from the RCS-2 and SDSS 
DR7 span redshift ranges of $0.2 \lesssim$~z~$\lesssim 1.2$ and 
$0.2 \lesssim$~z~$\lesssim 0.65$, respectively.

In order to identify strong lenses, each optically selected galaxy cluster 
is independently inspected by multiple experts and scored 
for evidence of strong lensing. Inspectors are also randomly made to re-score 
lines-of-sight that they have already scored, thereby providing data that will be used 
to quantify the consistency of our scoring, as well as our final completeness as a function 
of the averaged scores. We have also conducted exhaustive follow-up observations 
to quantify the purity of the resulting sample as a function of average score. 
A complete description of the SGAS and RCSGA samples are forthcoming (M.~D. Gladders 
et al., in preparation, and M.~B. Bayliss et al., in preparation, respectively). The 
giant arcs used in this paper are located around $97$ unique galaxy clusters that 
span the full range in photometric redshifts of the parent catalog of optically 
selected clusters described just above, with a median lens redshift of $\bar{z}_{l} = 
0.58$. In Figure \ref{lenses} we show the the probability distribution of foreground 
cluster lens redshifts for the giant arc sample.

The giant arc sample presented in this paper is an \emph{incomplete} subset of the 
full SDSS and RCS-2 giant arc samples.  In the paper we analyze photometric data for 
$105$ systems which constitute a large fraction of the final combined SGAS $+$ RCSGA 
samples. The criteria for inclusion in our analysis here are: 1) the giant arc 
was identified in a systematic search of $g-$band
imaging with depth matching the limits of the RCS-2 $g-$band survey data, and 2) that
we possess photometric $u-$band imaging of that arc. The $u-$band data is essential for 
classifying giant arcs based on well-established photmetric dropout criteria for distant 
blue galaxies \citep[e.g.,][]{steid1996a,steid1996b,lowenthal1997}.  All $u-$band observations 
presented in this paper were obtained at the 4.1m Southern Astrophysics Research (SOAR) 
Telescope located on Cerro Pachon in the Chilean Andes during the 2008B and 2009A semesters. 
In practice this means that $105$ arcs discussed in the paper represent those arcs which
had the appropriate $\alpha , \delta$ to be observable during the SOAR observing runs,
\emph{and} were discovered prior to the observing runs. The southerly location of SOAR
restricts us to arcs with $\delta \lesssim 20$ degrees, and because our systematic   
giant arc search is a process that is still on-going, there are many arcs which 
are now in the sample and have appropriate $\delta$ to be observable from SOAR 
but were not known at the time of the $u-$band imaging runs.

As discussed in the introduction, it is important to measure the properties for a large, 
unbiased sample of giant arcs. To that end we note that targets that were observed with 
SOAR and analyzed here were selected independent of their respective arc radii, 
R$_{arc}$. The arc radius is measured as the mean angular distance on the sky between 
a giant arc and the center of the foreground cluster lens, where the center is 
approximated as the center of the BCG. The sample analyzed here includes arcs spanning 
a large range in arc radius -- $3\arcsec \lesssim$ R$_{arc} \lesssim 55\arcsec$ -- and 
the arcs were identified in imaging data that is approximately $1.5-2$ magnitudes (point 
source) deeper than those used in \citet{bayliss2011a}, which was restricted to arcs that 
were visually identified in the SDSS DR7 imaging data.

Some of the giant arcs analyzed in this paper appear in the literature and have 
published spectroscopic redshifts. RCSGA J032727-132609 appears in \citet{wuyts2010} 
and \citet{rigby2011}; 
it is the most spectacular giant arc discovered to date in RCS-2 (and one of the most 
spectacular systems in the observable universe). RCSGA J152745+065219 is previously 
published as SGAS J152745+065219 \citep{koester2010,bayliss2011b}, and is located in 
a region where the RCS-2 and SDSS footprints overlap. SGAS 211119-011429 was first 
identified as a probably giant arc in \citet{hennawi2008}, and both  SGAS 095739+050928 
and SGAS 211119-011429 have spectroscopic redshifts published in \citep{bayliss2011b}. 
Lastly, the so-called ``Cosmic Eye'' \citep{smail2007} appears in our arc sample, as 
it is located near on the sky to a massive foreground galaxy cluster, though 
the arc itself is formed around an elliptical galaxy located behind the galaxy cluster. 
We include this arc in our sample because it appears in our visual inspection 
of cluster lines-of-sight, and the foreground cluster contributes significantly to 
the lensing \citep{smail2007}. Several other arcs analyzed in this paper have
been published as candidate but unconfirmed giant arcs or cluster lenses, including
SGAS 084647+044608, SGAS 085429+100819, SGAS 111504+164533, and RCSGA 004827+031114
\citep{wen2011}.

\begin{figure}
\centering
\includegraphics[scale=0.55]{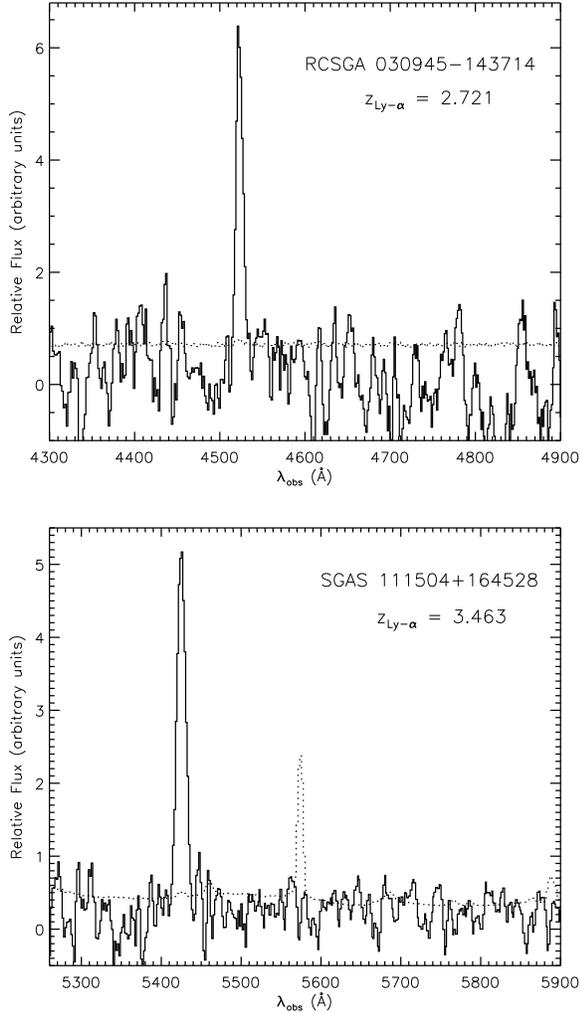}
\caption{\scriptsize{Optical spectra with strong Lyman-$\alpha$ emission from two arcs with optical
colors measured in this work that identify them as z $\geq 2.7$ galaxies. The spectroscopic
redshifts corroborate the photometric redshift assignment. In each panel the solid histogram
represents the measured flux per spectral pixel, and the dotted histogram is the error
array.
$Top:$ Optical spectrum of RCSGA 030945-143714 taken with the Goodman Spectrograph
on the SOAR 4.1m telescope.
$Bottom:$ Optical spectrum of SGAS 111504+164528 taken with DIS on the 3.5m telescope
at APO.}}
\label{newspectra}
\end{figure}

\begin{figure*}[t]
\centering
\includegraphics[scale=0.54]{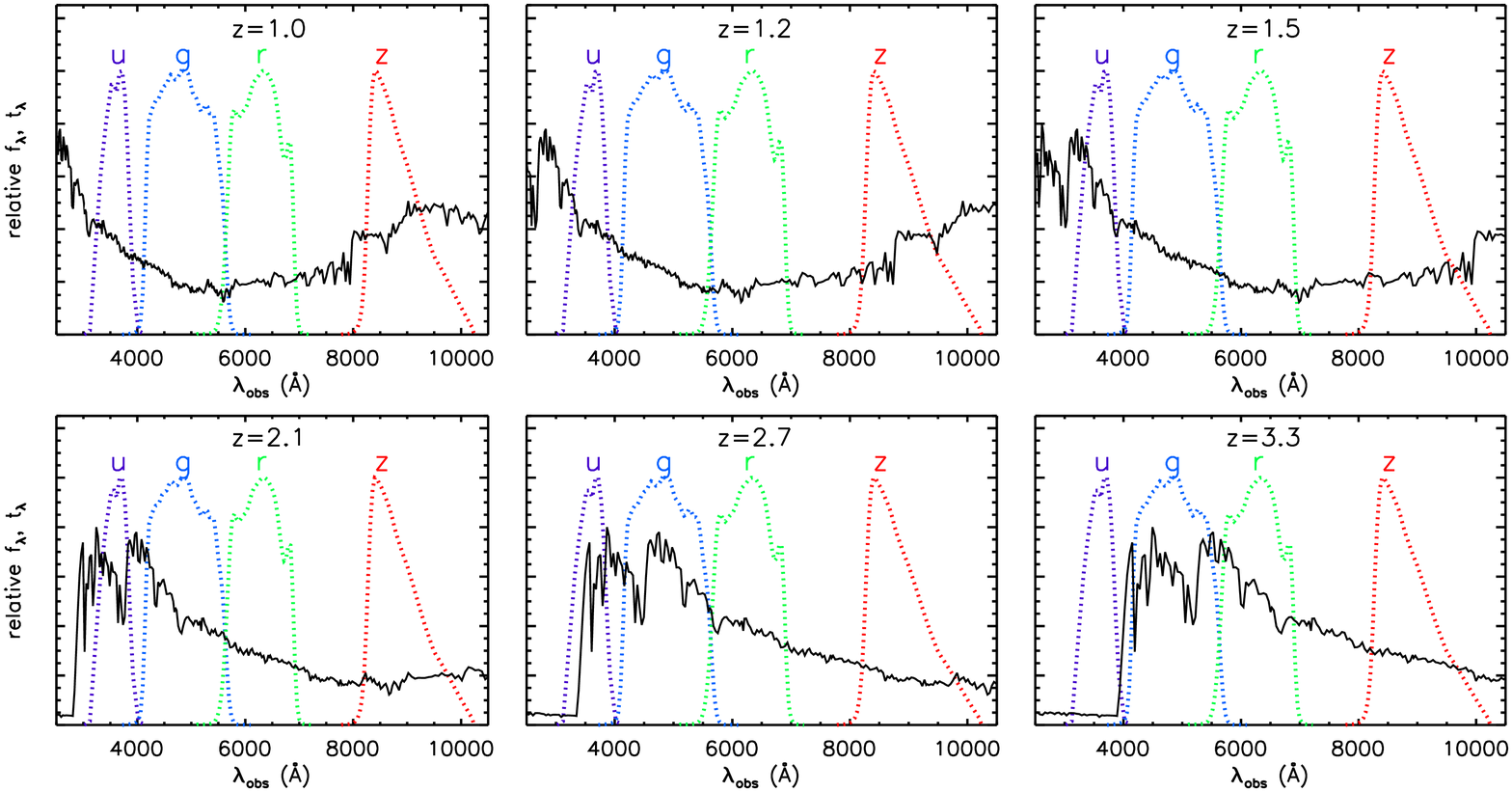}
\caption{\scriptsize{Similar to Figure $1$ in \citet{adelberger2004}, six panels show a synthetic
galaxy spectral energy distribution (SED) at a series of different redshifts,
with the $ugrz$ filter transmission curves over-plotted as colored dotted
lines. The ratio of galaxy fluxes measured in each of the different filters
can reliably identify galaxies within different redshift intervals based
on the location of strong spectral features, such as the Lyman-limit
at $912$\AA~(rest-frame), and the Balmer/$4000$\AA~break. These strong
spectral features appear generically in star-forming galaxies and can be
used to assign approximate redshifts for star-forming galaxies without
attempting to constrain their specific stellar populations and star-formation
histories.}}
\label{sed_filters}
\end{figure*}

\subsection{Follow-up $u-$band Imaging}
\label{sec:uband}

All $u-$band observations were conducted as part of NOAO programs 2008B-0400 
and 2009A-0414 (PI: M. Bayliss) on the 4.1m SOAR Telescope with the SOAR Optical 
Imager (SOI). SOI is a small mosaic of two $4096\times2048$ pixel CCDs filling a 
field of view of $5.25\times 5.25$ arcmin, with an unbinned pixel scale 
of $0.0727$\arcsec~pixel$^{-1}$. 
Observations of all but one of our targets were taken with the detector 
binned $2\times2$, where the remaining target was observed with the detector 
binned $4\times4$. Individual exposures varied between $120$s and $900$s 
and were dithered to cover the chip gap in the SOI mini-mosaic. Typical total 
integration times for individual targets range from $240$s to $3780$s, tuned to 
provide a minimum imaging depth complementary to the brightness of each visually 
selected giant arc in the RCS-2 $g-$band data, with a few systems of particular 
interest being imaged more deeply \citep[e.g.,][]{wuyts2010}.

Basic image reductions were performed using a combination of custom IDL code 
and the MSCRED IRAF\footnote{IRAF (Image Reduction and Analysis Facility) is distributed 
by the National Optical Astronomy Observatories, which are operated by AURA, Inc., 
under cooperative agreement with the National Science Foundation.} package. 
The custom IDL code was employed to remove time-variable detector-based noise 
structures that appeared in images, as well as for alignment of individual exposures. 
Flat-fielding, bias correction, and stacking were done using standard MSCRED routines. 
The resulting stacked $u-$band images were calibrated from stars in the SDSS DR7 
\citep{sdssdr7} that were observed  over a range of airmasses: $1.1 < Z < 1.7$. 
For each standard star observation, we solve the following equation:

$$m_{u} = m_{I,u} + c_{0} + c_{1}\times(Z - 1) + c_{2}\times(m_{u - m_{g}})$$

where $m_{u}$ and $m_{g}$ are the true apparent magnitudes of the 
standard in the $u-$ and $g-$bands, respectively, $m_{I,u}$ is the 
measured instrumental magnitude of the standard in a given $u-$band 
observation, and $Z$ is the airmass of a given observation. By solving 
the system of equations resulting from all of our standard observations 
we determine best-fit values for the photometric zero point $c_{0}$, the 
atmospheric extinction term $c_{1}$, and the photometric color term $c_{2}$ 
for our $u-$band observations. The average best-fit values for these terms 
in 2008B and 2009A observations is $c_{1}=0.51$ and $c_{2}=0.02$. The 
limiting source of error in the $u-$band 
photometric zero point determination comes from the fundamental uncertainty 
in the SDSS $u-$band calibration; $u-$band data is notoriously difficult to 
calibrate absolutely, a fact that has been well-documented in SDSS photometry 
\footnote{see http://www.sdss.org/dr7/start/aboutdr7.html}. From the intrinsic 
scatter in our standard star observations, and the published 
photometric characterization for the DR7 \citep{sdssdr7}, we adopt an upper limit 
on the uncertainty in our $u-$band photometry of $\pm 0.05$ mags. We also 
note that there is an established red light leak in the SDSS $u$ filter that 
causes slight biases in photometry for red objects, and we therefore restrict 
stars used in our zero point calibrations to types K and bluer.

\begin{deluxetable}{cl}
\tablecaption{Redshift Bins Definitions in Color-color Space\label{colortab}}
\tablewidth{0pt}
\tabletypesize{\tiny}
\tablehead{
\colhead{Redshift Bin} &
\colhead{Color Criteria } }
\startdata
$2.7 \leq z < 3.5$:  & ~~~~$g-r ~ \geq ~ -0.35$,  \\
~~     & ~~~~$g-r ~ \leq ~ 1.2$, \\
~~  & ~~~~$u-g ~ \geq ~ g-r + 1.0$  \\
~~~  &   ~~~  \\
\hline
~~~  &   ~~~  \\
$1.9 < z < 2.7$:  &  ~~~~$g-r ~ \geq ~ -0.35$,  \\
~~   &  ~~~~$g-r ~ \leq ~ 0.2(u-g) + 0.4$, \\
~~   &  ~~~~$u-g ~ \geq ~ g-r + 0.2$,  \\
~~   &  ~~~~$u-g ~ < ~ g-r + 1.0$  \\
~~~  &   ~~~  \\
\hline
~~~  &   ~~~  \\
$1.4 < z < 2.1$:  &  ~~~~$g-r ~ \geq ~ -0.35$,  \\
~~   &  ~~~~$g-r ~ \leq ~ 0.2(u-g) + 0.4$, \\
~~   &  ~~~~$u-g ~ \geq ~ g-r - 0.15$,  \\
~~   &  ~~~~$u-g ~ < ~ g-r + 0.2$  \\
~~~  &   ~~~  \\
\hline
~~~  &   ~~~  \\
$1.0 < z < 1.5$:   &  ~~~~$r-z ~ \geq ~ 0.8(g-r) + 0.3$,  \\
~~  &   ~~~~$g-r ~ < ~ 1.5$  \\
\enddata
\end{deluxetable}

\subsection{Survey Imaging of RCS Giant Arcs (RCSGA) Survey Objects}

The RCS-2 survey consists of $grz$ imaging over approximately $700$ deg$^{2}$
taken in queue mode with MegaCam at the 3.6m Canada-France-Hawaii Telescope
(CFHT) between 2005 and 2008. Individual RCS-2 pointings have single exposures
of $240$s, $480$s, and $360$s in each of the $grz$ filters, respectively.
Because these data are single pointings in each filter, the images contain
cosmic rays and chip defects which we remove manually when they occur near
one of our giant arcs. The RCS-2 survey data is fully calibrated
to the SDSS in all bands, which includes a color term describing the
difference between the CFHT MegaCam $g-$band and the original SDSS $g-$band.
The RCS-2 imaging is remarkably uniform in depth and seeing, making it an
ideal dataset for homogeneously selecting giant arcs. For further details on the
RCS-2 data we refer the reader to \citet{gilbank2011}, which describes the RCS-2
survey data in great detail.

\subsection{Optical Imaging of Sloan Giant Arcs Survey (SGAS) Objects}

The primary $g-$band imaging of SGAS systems was obtained at $2.5-4$m class telescopes 
over several years \citep[for examples and details of these observations see][]{hennawi2008}, 
and it is these data from which the SGAS arcs analyzed here are systematically identified. 
However, there exists for some of the SGAS giant arcs analyzed in this paper a 
variety of deeper imaging which we use for photometric measurements where it is 
available. A subset of the SGAS objects in this paper were also observed in the 
$g-$ and $r-$bands with the $8$m Gemini South Telescope with the GMOS instrument 
\citep{bayliss2011b} and/or the $8.2$m Subaru Telescope 
\citep{oguri2009,oguri2011} as a part of a larger 
program designed to collect extensive follow-up observations for a large 
subset of the full SGAS sample. We use the best available data for all photometric 
measurements presented in the paper in order to achieve the highest quality 
measurements possible. Four of the SGAS objects analyzed here have no deep $r-$band 
imaging available beyond the publicly available SDSS survey imaging, and so we 
use these data where necessary. All of the arcs, though confirmed in deeper $g-$band 
imaging, are sufficiently bright as to be well-detected in the SDSS $r-$band imaging 
data.

\subsection{New Spectroscopy of Individual Giant Arcs}
\label{sec:newspec}

\begin{figure*}[t]
\centering
\includegraphics[scale=0.7]{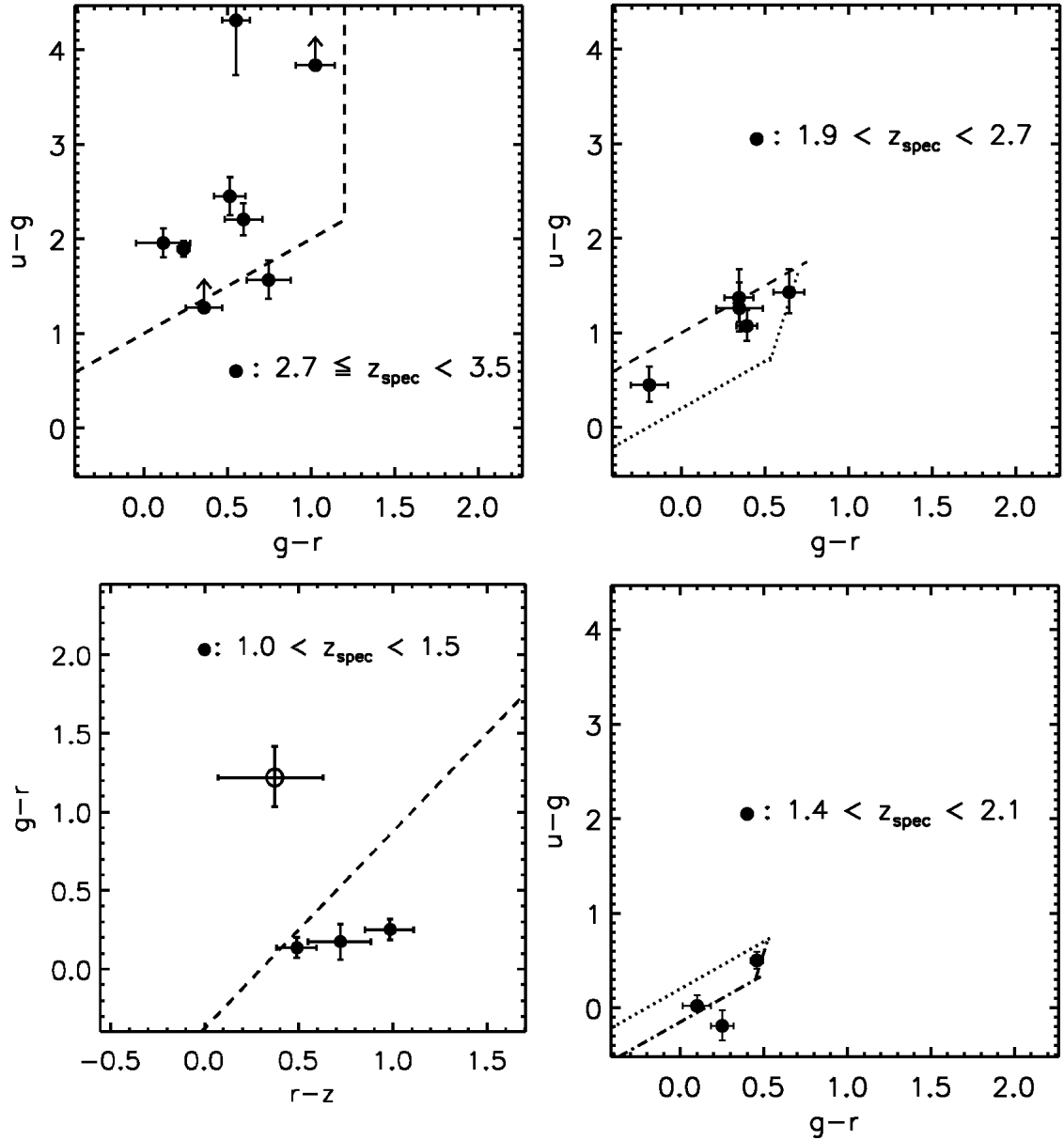}
\caption{\scriptsize{Giant arcs with known spectroscopic redshift are plotted in
their corresponding color-color space. Each panel plots arcs which match a
different redshift bin in color-color space. The bins, beginning in the upper
left and proceeding clockwise are, $2.7 \leq $~z~$ < 3.5$, $1.4 < $~z~$ < 2.7$,
$1.4 < $~z~$ < 2.1$, and $1.0 < $~z~$ < 1.5$. There is some small overlap between some
of the neighboring bins, and some arcs are therefore plotted in multiple panels.
Arcs with known redshifts agree very well with the corresponding regions in
color-color space, with $1/19$ lying approximately $1.8\sigma$ outside of the
expected region and several others lying $1\sigma$ or so outside of the
appropriate regions. One arc with a spectroscopic redshift, z$_{spec} < 1$ is
plotted as an open circle in the lower left panel, and as expected it falls
outside of the color-color region corresponding to $1.0 < $~z~$ < 1.5$.}}
\label{speccheck}
\end{figure*}

In this paper we also present new spectroscopic observations of two arcs in our photometric 
sample. RCSGA 030945-143714  and RCSGA 030945-143717 were observed with the Goodman 
Spectrograph \citep{clemens2004,crain2004} on November 1, 2008 at the end of the first 
SOAR $u-$band observing run mentioned above (NOAO 2008B-0400). The Goodman Spectrograph is 
an imaging spectrograph with multi-object capabilities designed to use Volume Phase 
Holographic (VPH) transmission gratings and optimized for throughput in the wavelength range, 
$\sim3200-8000$\AA, especially in the blue. We selected the target for spectroscopy 
from the full $u-$band imaging target list based on the presence of two bright arcs in the 
field (RCSGA 030945-143714 and RCSGA 030945-143717) that could be observed simultaneously 
by positioning the slit approximately $40$ degrees East of North so as to place both arcs 
within the slit. The spectrograph was configured with a $1.03\arcsec$ wide longslit mask, 
the KOSI\_600 grating, and camera \& grating angles of $20$ \& $10$ degrees, respectively. 
The detector was binned $2\times2$, resulting in a spatial scale along the slit of 
$0.3\arcsec$ pix$^{-1}$ and a mean dispersion of $1.31$\AA pix$^{-1}$ and a spectral 
full width at half max (FWHM) of $4.2$\AA. This mode provides a central wavelength of 
$5696$\AA~ and wavelength coverage over the range, $\Delta \lambda = 4200-7000$\AA.

Science observations of these targets consist of $4\times1800$s exposures, with quartz 
lamp flatfields and HeAr lamp calibration exposures bracketing individual science 
exposures. The data were reduced, extracted, and stacked using custom IDL scripts that 
incorporate procedures from the XIDL \footnote{http://www.ucolick.org/$\sim$xavier/IDL/index.html} 
software package. At the position of RCSGA 030945-143714 along the slit there is a single 
mildly asymmetric emission line feature is evident in all four individual exposures at $4523$\AA, 
which we determine to be Lyman-$\alpha~\lambda1216$\AA~ at a redshift, z$_{arc} = 2.721$. Other 
plausible interpretations for the emission feature (i.e. nebular emission lines at lower redshift) 
are ruled out by the absence of additional lines in the wavelength range redward of the 
emission feature ($\Delta \lambda = 4600-7000$\AA). If the emission line was, in fact, 
O[II]~$\lambda3727$\AA~ or H-$\beta~\lambda4341$\AA~ then we would expect to also see 
one or more of H-$\beta~\lambda4862$\AA, O[III]~$\lambda4960,5007$\AA~ or 
H-$\alpha~\lambda6563$\AA) in the wavelength range covered by our observations. We 
can also note that in our final photometric analysis RCSGA 030945-143714 falls into 
the appropriate region of the $u-g$ vs $g-r$ space for a galaxy with redshift z: 
$2.7 \leq z < 3.5$. At the location on the slit corresponding to RCSGA 030945-143717 
there is some weak continuum signal spanning the highest-throughput wavelength range 
for the setup, but nothing sufficient to measure a redshift.

\begin{deluxetable*}{ccc}[t]
\tablecaption{Giant Arcs With Published Redshifts\label{spectable}}
\tablewidth{0pt}
\tabletypesize{\tiny}
\tablehead{
\colhead{Giant Arc} &
\colhead{z$_{spec}$ } &
\colhead{Reference} }
\startdata
RCSGA 030945-143714 & $2.719$  & this work  \\
RCSGA 032727-132609 & $1.704$  & \citet{rigby2011}  \\
RCSGA 213513-010143 & $3.07$   & ``Cosmic Eye'' \citep{smail2007} \\
SGAS 095739+050929  & $1.820$  & \citet{bayliss2011b,diehl2009}  \\
SGAS 111504+164528  & $3.463$  & this work  \\
SGAS 152745+065219  & $2.760$  & \citet{koester2010,bayliss2011b} \\
SGAS 211119-011432  & $2.858$  & \citet{bayliss2011b}  \\
\enddata
\end{deluxetable*}

A second giant arc from our sample, SGAS 111504+164528, was also observed with the Dual 
Imaging Spectrograph (DIS) on the 3.5m Astrophysical Research Consortium (ARC) Telescope 
at Apache Point Observatory in New Mexico on the night of May 23, 2011. These observations 
were conducted with DIS in $1.5\arcsec$ longslit mode using the B400/R300 gratings in the 
``low-res'' configuration. Science exposures consisted of $2x2400s$, with quartz lamp 
flatfields and HeNeAr lamp calibrations bracketing the science exposures. Blue-side 
spectra have a mean dispersion of $\delta \lambda = 1.83$\AA pix$^{-1}$ and a spectral 
FWHM of $5.3$\AA, and red-side spectra have mean dispersion, $\delta \lambda = 2.31$\AA 
pix$^{-1}$ and a spectral FWHM of $6.4$\AA. The blue and red channels combine to 
provide full optical wavelength coverage over the range, $\Delta \lambda = 3800-9800$\AA.
These data were reduced and analyzed in the same fashion as the Goodman spectroscopy described 
above using custom IDL scripts incorporating procedures from the XIDL package. 

DIS uses a dichroic splitting optic centered at $\sim5500$\AA~, and we observe a single strong 
emission line in both the blue- and red-side spectra corresponding to the slit position of 
SGAS 111504+164528, centered at $5425$\AA. Similar to the case of RCSGA 030945-143714 
described above, we conclude that this feature is Lyman-$\alpha~\lambda1216$\AA~ at a 
redshift, z$_{arc} = 3.463$, based on the lack of other visible emission features over the 
observed wavelength range $\Delta \lambda = 3800-9800$\AA. Similar to the case for 
RCSGA 030945-143714 above, the SGAS 111504+164528 has $u-g$ vs $g-r$ colors that also 
identify it as a galaxy in the redshift range, $2.7 \leq z < 3.5$. Extracted spectra for 
RCSGA 030945-143714 and SGAS 111504+164528 are shown in Figure~\ref{newspectra}.

\section{Measurements and Methodology}
\label{sec:methods}

\subsection{Optical Selection of Star Forming Galaxies}
\label{sec:colorcuts}

Historically, Lyman Break Galaxies (LBGs) are selected by examining wide-band 
photometric data and identifying the redshifted `Lyman limit' continuum break. 
This strong spectral break appears at 912\AA~ in the rest frame 
\citep{steid1996a,steid1996b,lowenthal1997}, and moves 
redward in the rest-frame with increasing redshift -- approaching 1216\AA~ -- due 
to Lyman-$\alpha$ forest \citep{steid1987,rauch1998} absorption by intergalactic 
neutral hydrogen. Surveys for LBGs are efficient for collecting statistical samples 
of high-redshift galaxies because it is difficult for galaxies at lower redshifts to 
mimic the sudden and extreme spectral break of the Lyman Limit. \citet{steidel2003} 
published the comprehensive results of a systematic search for z $\gtrsim2.7$ galaxies,
and this work was extended to probabilistically sort star-forming galaxies into bins 
at lower redshift intervals by identifying the color-evolution that occurs as additional 
weaker spectral features redward of the Lyman Limit redshift through a set of optical 
filters \citep{steidel2004,adelberger2004}. The ratio of galaxy fluxes -- or colors -- 
measured in different broadband near-UV, optical, and near-infrared filters can reliably 
identify galaxies within different redshift intervals based on the location of 
strong spectral features, such as the Lyman-limit at $912$\AA~(rest-frame), and 
the Balmer/$4000$\AA~break (see Figure~\ref{sed_filters}). These strong spectral 
features appear generically in star-forming galaxies and can be used to assign 
approximate redshifts for star-forming galaxies without attempting to constrain 
their specific stellar populations and star-formation histories. We use this 
color selection technique and use it to identify the fraction of a complete 
sample of $105$ giant arcs which fall into four broad redshift bins at z $>1$ 
(and by exclusion, giant arcs at z $<1$ are also identified).

In order to define photometric dropouts we follow the methodology of 
\citet{steidel2003}, \citet{steidel2004}, and \citet{adelberger2004} to define regions 
in color-color space that correspond to distinct redshift bins. The available photometric 
data is in a set of filters that are similar to those used in 
\citet{steidel2003}, \citet{steidel2004}, and \citet{adelberger2004}, with the only significant 
differences being an $r-$band filter with somewhat different characteristics. We use synthetic 
galaxy spectra from \citet{bruzchar2003} with identical stellar age and star formation history 
properties as are used by \citet{steidel2003}, \citet{steidel2004}, and \citet{adelberger2004}  
and step these model spectra through redshift steps of $\delta z=0.01$ from z $=0.5$ to z $=4.0$ 
and measure the resulting synthetic colors. These simulated galaxy colors as a function of 
redshift are essentially identical to those appearing in \citet{steidel2003}, 
\citet{steidel2004}, and \citet{adelberger2004} with slight differences of $\sim0.1-0.15$ 
mags in colors which include the $r-$band filter. The final color cut criteria that we apply 
to our photometry defines four primary redshift bins, all of which are summarized in 
Table~\ref{colortab}.

Arcs that do not satisfy any of the criteria in Table~\ref{colortab} are likely 
to have redshifts that are either greater than $3.5$ or less than $1.0$. In 
principle we might be able to identify very high redshift arcs in our sample 
(e.g. z $>3.5$) as $g-$band dropouts using logic that is analogous to the 
original $u-$band dropout definitions from \citet{steidel2003}. 
However, we are limited by the lack of deep imaging redward of the 
$r-$band. In any event, arcs that are very red in $g-r$ are rare 
($N_{g-r > 1.2} = 6$), and all of these objects either have $r-z > 0.9$, or 
are detected in $z$ at less than $4\sigma$. This leads us to conclude that arcs which 
are red in $g-r$ are likely to be intrinsically red galaxies at z $<1$. That stated, it 
is still possible that our arc sample contains one or two very high redshift 
$g-$band dropout galaxies, and we account for this possibility in the final 
uncertainties on our constraints for the giant arc redshift distribution.

\subsection{Aperture Photometry of Giant Arcs}
\label{sec:appphot}

We measure four- and three-band photometry for a full sample of $105$ giant arcs
from the RCSGA and SGAS samples, respectively. All objects have measurements in
$ugr$, and objects located in the RCS-2 fields have $z-$band measurements as
well. Photometric measurements are conducted using a custom IDL pipeline that
allows us to construct photometric apertures which follow the ridge-lines of
giant arcs and match the highly variable structures of these objects on the
sky, following the procedure used by \citet{wuyts2010}. We draw apertures by
manually defining a ridge-line along each giant arc, and convolve the ridge-line
with the point spread function (PSF) of each image to create a series of apertures
of increasing radius, following isophotes of the convolution. Arcs are typically
only resolved along a single axis, because the lensing magnification acts almost
exclusively along the tangential direction (relative to the center of the
potential of the foreground lens). We therefore measure photometry in each image
out to an aperture radius equal to $2.0\times$FWHM, where the FWHM describes the 
PSF measured from reference stars in each image. These
apertures allow us to measure an equivalent region on the sky for images with
different PSFs. All magnitudes are aperture corrected to a radius of $6\arcsec$
using the curve of growth for reference stars in each image.

In cases where the flux for a given arc within its $2.0\times$FWHM defined aperture 
is less than $2\sigma_{app}$, where $\sigma_{app}$ is the noise measured within 
the aperture, we measure a limiting magnitude for that object in that filter. 
Limiting magnitudes are measured as the magnitude which corresponds to 
$2\sigma_{app}$, and this limiting magnitude is aperture corrected to the same 
$6\arcsec$ radius as above. Quoted limiting magnitudes are $95\%$ confidence upper 
limits on the flux from the corresponding arc within its aperture. All photometry is 
corrected for galactic extinction \citep{schlegel1998}.

The ridge-line apertures used for this analysis are not necessarily intended to enclose 
all of the observable flux from each lensed background source. Rather, we draw
apertures along the largest contiguous image or arc of each lensed galaxy, and make 
no attempt to add flux from multiple images/arcs, or faint arc tails that appear to 
originate or extend away from the same source. Our measurements are conducted with 
robust color estimates as our primary goal, and our ground-based imaging typically 
does not have either the depth or image quality to unambiguously identify complete 
arc families in the same way as can be achieved with adaptive optics or $HST$ quality 
imaging.

We also note that variations in color that can be observed in different lensed images 
of a single background source may provide a distorted \citep[e.g. cB58;][]{williams1996}
representation of the intrinsic source. That stated, color gradients for star
forming galaxies tend to be small -- on the order of $\sim0.1$ mags \citep{suh2010}. 
Color distortions due to magnification gradients across the surface of the lensed 
sources should therefore be well within the color variations accounted for in the 
definition of the regions in color-color space, which are intended to apply to 
star forming galaxies with a variety of colors and there is no reason to expect 
catastrophic outliers (i.e. giant arcs with apparent colors that differ dramatically 
from their true intrinsic colors) to be a significant problem for our analysis. 

Approximately $1/3$rd of the arcs analyzed here are located nearby on the sky to 
other, presumably foreground, galaxies. For these objects there is light from a projected 
source, typically the brightest cluster galaxy (BCG) or a cluster member galaxy in the 
foreground lens, that overlaps with the $2.0\times$FWHM aperture that we use to measure 
giant arc arc magnitudes. We have used the publicly available GALFIT package 
\citep{peng2010} to fit Sersic profiles to the nearby galaxies and subtract the 
resulting model fluxes. Galaxy light profiles are fit in a filter where the galaxy 
signal-to-noise (S/N) is high, and that profile is then saved and scaled in brightness to 
match the galaxy light in the other imaging bands. A single Sersic profile is almost 
always sufficient to the task for this work because we are generally concerned with 
subtracting off the extended stellar light profile of BCGs, and we are not concerned 
with accurately modeling the detailed structure in the BCG core which has no impact 
on the arc photometry. Occasionally there are very faint intervening galaxies that 
lack sufficient S/N to achieve a robust GALFIT model, and we remove these 
galaxies by manually masking the galaxy pixels. 

\begin{figure}
\centering
\includegraphics[scale=0.6]{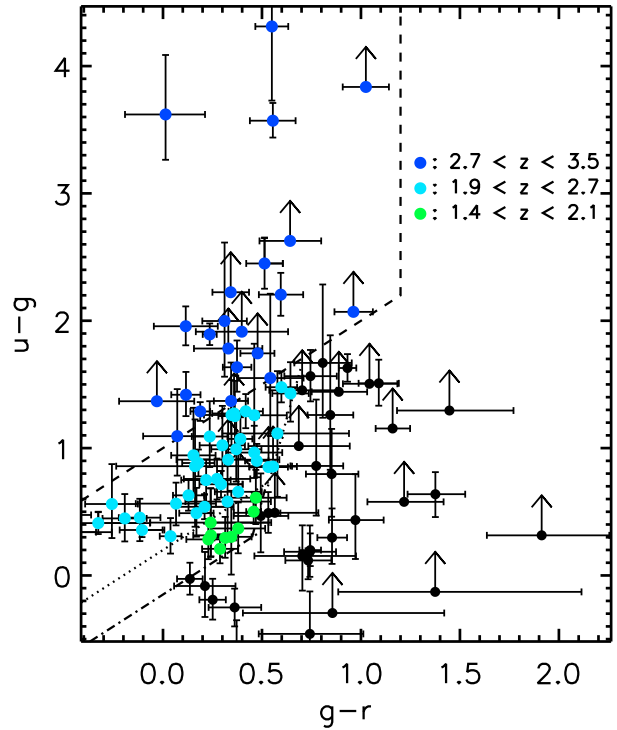}
\caption{\scriptsize{Photometric colors, $u-g$ vs $g-r$, for all $105$ giant
arcs that have $u-$band data from SOAR. Three regions in the color-color
space are defined by the dashed, dotted, and dot-dashed lines corresponding
to redshift ranges, $2.7 \leq $~z~$ < 3.5$, $1.9 < $~z~$ < 2.7$, and $1.4 < $~z~$ < 2.1$,
respectively. Arcs that fall into each respective bin are plotted as blue,
cyan, and green points, respectively, while arcs not falling into any of the
aforementioned redshift bins are plotted in black; by exclusion these objects
have redshifts $<1.4$.}}
\label{ug_gr_plot}
\end{figure}

\begin{figure}
\centering
\includegraphics[scale=0.6]{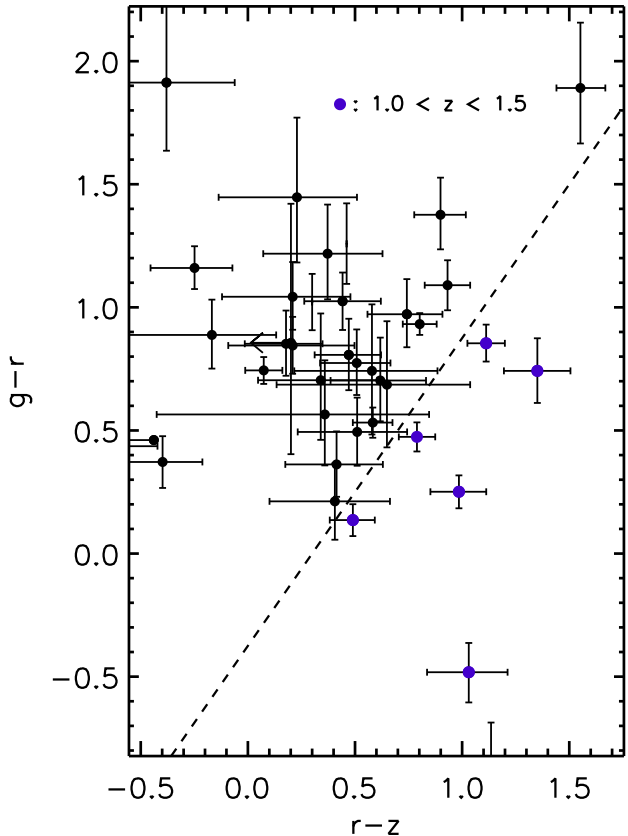}
\caption{\scriptsize{Colors, $g-r$ vs $r-z$, for $96$ arcs
discovered in the RCS-2 survey area. The dashed line indicates the division
in the color-color space that identifies typical star forming galaxies at
$1.0 < $~z~$ < 1.5$ (these galaxies are located to the right/below the dashed
line). Arcs satisfying this criteria are plotted in purple, while other
arcs are plotted in black. Several arcs, including many of those with marginal
or poor $z-$band photometry fall outside of the color-color region shown in
the plot.}}
\label{gr_rz_plot}
\end{figure}

\subsection{Verification of Color-Based Redshift Assignment}
\label{sec:spectest}

We can directly test the validity of applying the \citet{steidel2003}, \citet{steidel2004}, 
and \citet{adelberger2004} color criteria for rough redshift assignment to our sample of 
giant arcs by comparing spectroscopic redshifts for individual giant arcs against their
color-color space redshift designation. All photometric measurements were made
blind to any prior redshift knowledge for our giant arc sample, and after the
fact we identify $19$ giant arcs in our sample that have known, high-confidence
spectroscopic redshifts. Five of these giant arc redshifts are in the literature,
while the remaining $14$ are the result of new observations. Two of these new
redshifts are presented here (see Section 2.5), and the remaining $13$ will
appear in forthcoming publications: one in M.~D. Gladders et al. 2011, in preparation
and $12$ in M. Carrasco et al. 2012, in preparation). The authors of these papers
have generously granted us access to the unpublished spectroscopic redshifts for the
purpose of verifying the color-color redshift designation method used in this
paper. The seven giant arcs with public redshift information (five published plus
two new arcs presented here) are summarized in Table~\ref{spectable}.

In order to verify the success of the color-color based methodology
when applied to our giant arc sample, we populate the relevant color-color spaces with
measured colors for arcs of known redshift in our sample and compare their location to
the regions defined based on the methods first published in \citet{steidel2003}, 
\citet{steidel2004}, and \citet{adelberger2004}. The resulting $u-g$ vs $g-r$ and
$g-r$ vs $r-z$ plots are shown in Figure~\ref{speccheck}, and provide excellent 
validation of the use of cuts in color-color space as a tool for estimating the 
redshifts of our giant arc sample. We can perform an additional sanity check by 
comparing the redshift bin assignments for the giant arcs against photometric 
redshifts for the foreground cluster lenses that are measured as a part of the 
red-sequence cluster finding algorithm. Indeed, we find that the giant arcs which 
are sorted into the z $\leq1$ redshift bin are not formed by clusters that have 
z$_{phot} \gtrsim0.8$, which is encouraging given that the geometry of such systems 
makes the occurrence of strong lensing difficult at best (and non-physical at worst). 
A broad examination of the relationship between z$_{lens}$/z$_{arc}$ pairs yields no 
significant correlation.
 
\subsection{Redshift Constraints On $105$ Giant Arcs}
\label{sec:constraints}

Giant arc $u-g$ vs $g-r$ and $g-r$ vs $r-z$ colors are shown in Figure~\ref{ug_gr_plot} and 
Figure~\ref{gr_rz_plot}, respectively. Individual giant arc color measurements are color-coded 
in the figures according to the redshift bins that they fall into. The $g-r$ vs $r-z$ color 
space in Figure~\ref{gr_rz_plot} is populated using only RCSGA survey objects, because we 
lack $z-$band imaging of sufficient depth for our SGAS arcs. RCSGA arcs, however, 
comprise $>90\%$ of our full giant arc sample, and we can also note that seven of the 
nine SGAS arcs fall into redshift bins based on their $u-g$ vs $g-r$ colors, so that 
the lack of $z-$band data for the SGAS sample only prevents us from assigning two 
systems to one of the z $<1.0$ or $1.0 < $~z~$ < 1.5$ bins, which has a negligible affect 
on our ability to characterize the redshift distribution of the sample as a whole. 

\begin{deluxetable}{cc}
\tablecaption{Fraction of Arcs Satisfying Different Color Criteria\label{redshift_frac}}
\tablewidth{0pt}
\tabletypesize{\tiny}
\tablehead{
\colhead{Redshift Bin} &
\colhead{Fraction of Arcs } }
\startdata
$2.7 \leq $ z$_{arc}$  & $0.21^{+0.04}_{-0.02} $  \\
$1.9 < $ z$_{arc} < 2.7$ & $0.33^{+0.03}_{-0.03}$  \\
$1.4 < $ z$_{arc} < 2.1$  & $0.09^{+0.04}_{-0.02}$  \\
$1.0 < $ z$_{arc} < 1.5$  & $0.08^{+0.04}_{-0.02}$  \\
z$_{arc} < 1.0$\tablenotemark{a} & $0.29^{+0.03}_{-0.05}$ \\
\enddata
\tablenotetext{a}{~Fraction of arcs with z $<1.0$ is measured by exclusion, i.e. giant
arcs with colors not meeting the other criteria covering the redshift range
$1.0 < z \lesssim 3.5$.}
\end{deluxetable}

The resulting constraints on the fraction of arcs falling within each redshift bin are 
shown in Table~\ref{redshift_frac}. We estimate the errors on the fractions of giant 
arcs meeting various redshift criteria by generating $10^{4}$ Monte Carlo realizations of 
our data, where each realization is a simulated dataset that is generated by assuming 
that each photometric measurement is gaussian distributed about the measured value, 
with standard deviations equal to the $1\sigma$ errors on the measurements. Some 
photometric measurements have asymmetric errors, and for these cases the simulated data 
accounts for the asymmetry in the errors. We make no attempt to recover a total probability 
distribution, $dp_{\rm arc}/dz$, because doing so requires detailed knowledge of the 
probability distribution of redshifts within each bin in color-color space, which 
we do not have. Instead we simply report the total cumulative fraction of giant arcs 
which have redshifts, z$_{arc}$, greater than various discrete cuts. The resulting 
cumulative redshift fractions are $71^{+5}_{-4}~\%$ at z $\geq1.0$, $64^{+6}_{-4}~\%$ at 
z $\geq1.4$, $56^{+5}_{-4}~\%$ at z $\geq1.9$, and $21^{+4}_{-2}~\%$ at z $\geq2.7$. The 
fraction of arcs with z$\geq1.4$ and z$\geq1.9$ requires making some assumptions about 
the probability distribution within each redshift bin, because some of the neighboring bins 
have regions of overlap in redshift. For example, in order to compute the fraction of giant 
arcs with z$\geq1.4$, we first account for all arcs that have colors identifying them as 
galaxies at $2.7 \leq $~z$_{arc} < 3.5$, $1.9 < $~z$_{arc} < 2.7$, and $1.4 < $~z$_{arc} < 2.1$. 
We then must estimate the fraction of arcs within the redshift range, 
$1.0 < $~z$_{arc} < 1.5$, that we expect to have z$_{arc} > 1.4$. For these reported 
values we assume that giant arc redshifts within a bin are roughly gaussian distributed within 
that bin with a standard deviation that is based on the results of spectroscopic follow-up 
for galaxies falling within these same color-based redshift bins in \citet{adelberger2004}. 

\begin{figure*}[t]
\centering
\includegraphics[scale=0.7]{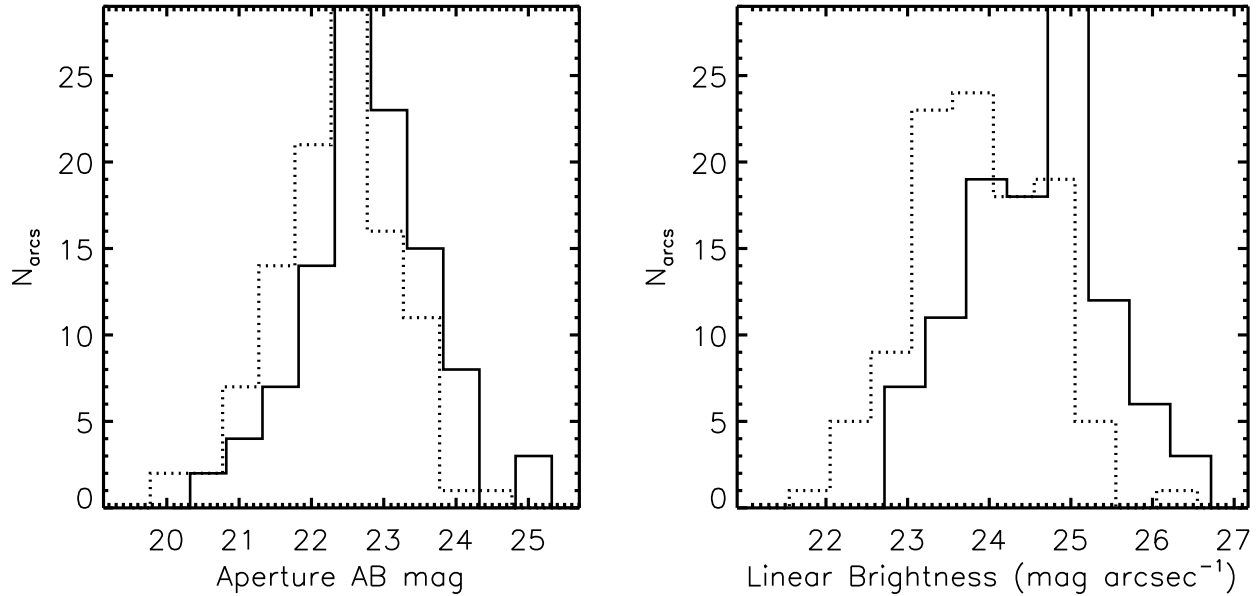}
\caption{\scriptsize{Distribution of magnitudes for all giant arcs
presented in this paper. $Left:$ Integrated magnitudes for
all arcs; the solid line histogram shows $g-band$ magnitudes,
and the dashed line histogram shows $r-$band magnitudes.
$Right:$ Linear magnitudes for the entire arc sample,
where the solid and dashed line histograms again
correspond to the $g-$ and $r-$band, respectively.}}
\label{arcsample}
\end{figure*}

\subsection{Giant Arc Linear Brightness Measurements}

In addition to the optical colors of our giant arc sample, we also measure the linear 
brightnesses for each arc in the $g-$ and $r-$bands. The linear brightness is computed 
for a given arc by dividing the total aperture magnitude for each arc by the length of 
the aperture, in arcseconds. This quantity is a very useful observable for giant arcs 
with ground-based imaging data, where the arcs are typically unresolved in the radial 
direction (with respect to the center of the foreground lensing potential). So long as 
an arc is resolved along the tangential direction (again, with respect to the center of 
the foreground lensing potential) -- a condition that is essentially always satisfied -- 
then the linear brightness is an observable that can be extracted from the data. The  
linear brightness has the appealing property of being independent of the PSF of the 
seeing of a particular image, and it is an observable quantity that can be easily 
produced for simulated giant arcs by convolving simulated imaging data with a smoothing 
kernel that is chosen to match the image quality of real imaging data. 

Figure~\ref{arcsample} shows the distribution of both integrated and linear magnitudes 
for our complete arc sample in both the $g-$ and $r-$bands. The measured distribution of 
linear magnitudes for a statistical sample of giant arcs is an observable which, like the 
arc redshift distribution, informs predictions of giant arc statistics by providing 
direct constraints on the photometric properties of simulated giant arcs. Specifically, 
the linear magnitudes of arcs should depend on some combination of two factors: 1) the 
typical magnifications for the lens-source systems, and 2) the intrinsic surface 
brightness function of the lensed background sources. Simulated arc samples must be able 
to reproduce the observed linear magnitude distribution in order to ensure that they are 
providing a robust simulation of the physical universe.

\section{Discussion and Summary}
\label{sec:discussion}

\subsection{A Census of Physical Factors Relevant to Predictions for Giant Arcs}

As we summarized in the Introduction, there has been no shortage of ideas for physical 
effects which, if unaccounted for in cosmological simulations, can explain varying degrees 
of discord between giant arc statistics predictions and observations. One such factor that 
has been directly addressed in the literature is the impact that baryons can 
have on the strong lensing cross-section for galaxy clusters 
\citep{puchwein2005,rozo2008,wambs2008}. \citet{rozo2008} find that 
the inclusion of baryonic cooling results in significant steepening of the central density 
profile relative to what is seen in dark matter only simulations and estimate that these 
processes can increase giant arc abundances by as much as a factor of a few, and 
\citet{wambs2008} find more modest increases of $\sim25$\%. However, the 
real-world impact is very likely to be smaller than measured by these authors due to 
the well-established ``over-cooling'' problem in simulations \citep[e.g.,][]{balogh2001}. 
In fact, more recent studies of strong lensing efficiencies in simulations that include 
gas physics find that the inclusion of feedback processes serves to mitigate the increase 
in strong lensing efficiency that result from including baryonic cooling physics 
\citep{mead2010}.

One key property of massive halos that does have a large impact on their ensemble 
lensing efficiencies is the triaxiality of their matter distributions. Halos identified 
in dark matter simulations are known to have triaxial shapes 
\citep{kasun2005,allgood2006}, 
and the triaxiality can cause large orientation-dependent variations in the strong lensing 
cross-section for individual halos \citep{oguri2003,dalal2004,hennawi2007,menegh2010}. 
This effect should be well-accounted for in predictions of giant arc counts using large 
volume simulations \citep[e.g.,][]{hennawi2007,menegh2010}. We must also point out, 
however, that the triaxiality of the matter distribution in the cores of clusters in 
simulations is sensitive to the presence/absence of gas physics. Baryonic cooling 
processes cause the shape of the density profile in cluster cores to become much 
less triaxial than in pure dark matter simulations \citep{kazan2004,rozo2008}, 
which could mitigate the importance of triaxiality and the magnitude of an 
orientation bias.

Simulations that account for the presence of baryonic matter by populating dark matter 
simulated halos with evolved mock galaxies may also be able to increase the number 
of expected giant arcs, but different studies claim varying increases in the 
efficiency of giant arc production. Dark matter only simulations with galaxies 
``painted on'' after the fact cannot account for the detailed effects that baryons 
have on the shape and steepness of the matter density profile, but the inclusion 
of a large concentration of baryonic matter in the cores of simulated halos does 
produce an increase in the strong lensing cross-section. The size of the impact 
that painted on galaxies have on giant arc counts is claimed to be as small as 
$\lesssim25\%$ \citep{flores2000,menegh2000,menegh2003}, and as large as a factor 
of $2$ \citep{hilbert2008}. Though the precise magnitude of the effect of 
including galaxies in ray-tracing simulations of giant arcs varies in the literature, there 
is no evidence that it approaches the factors of $\sim10$ necessary to account for the 
difference between current arc counts and the predictions by \citet{bartelmann1998}. 

The impact of dark matter substructure in cluster-scale halos has also been suggested as a 
potentially important factor \citep{flores2000,menegh2000,menegh2003}, but work done to 
quantify the contribution of substructure to lensing efficiency suggests that it is 
negligible \citep{hennawi2007}. Other recent work indicates that the efficiency for giant 
arc production of individual clusters seems to be in good agreement when comparing the 
number of giant arcs produced \emph{per} \emph{cluster} in $HST$ imaging of X-ray selected 
cluster samples against mock images created from ray tracing modern simulated massive 
clusters \citep{horesh2005,horesh2011}. 

\citet{horesh2011} use the Millennium Simulation \citep{springel2005} with galaxies included 
via a semi-analytic prescription, so that the galaxy contribution to the lensing efficiency 
of individual clusters is accounted for. Notably, this work \emph{does not} address the 
question of the overall abundance of giant arcs -- e.g. giant arc counts over a 
well-quantified fraction of the sky imaged to a well-defined depth -- but it does isolate 
the question of whether there is a systematic discrepancy between the strong lensing 
efficiencies of individual simulated vs. real clusters. As summarized above, a study of the 
global increase in giant arc counts in the Millennium Simulation, including galaxies, 
suggests a factor of $2$ increase over the earlier work of \citet{bartelmann1998}. The 
\citet{horesh2011} result therefore imply that the apparent discrepancy between the real and 
predicted giant arc abundances is primarily due to factors other than those having to do with 
the internal structure of galaxy clusters (e.g., contributions to the project surface mass 
density from abnormally high concentrations, substructure, or cluster galaxies).

With clear motivation now to focus on factors beyond the internal properties of simulated 
clusters, there are several potential effects on which to focus our attention. Two 
additional ideas that have been explored in the literature are the contribution to cluster 
lensing efficiency from large scale structure (e.g. filaments and uncorrelated halos) along 
the line of sight looking out toward massive clusters \citep{wambs2005,hilbert2007,puchwein2009}, 
and short timescale increases in the lensing cross-sections of individual clusters due to 
major mergers \citep{torri2004}. Studies using the Millennium Simulation conclude that the 
resulting increase in the strong lensing efficiency from these unrelated, projected density 
effects is in the $\sim30\%$ range, and are therefore not sufficient to significantly ease 
the tension between giant counts observed and predicted. Observational tests looking for 
projected line-of-sight structure toward strong lenses \citep{faure2009,fassnacht2011} find 
no evidence of an increase in the amount of uncorrelated structure near lenses vs. the field, 
in agreement with the simulation-based results.

With regard to mergers, \citet{torri2004} find an increase of nearly an order of magnitude 
in an individual cluster's strong lensing cross-section 
during a merger, the short timescale of the boosted cross-section and the infrequency of 
major mergers in cluster-scale halos make it difficult for mergers to contribute 
substantially to the integrated efficiency of galaxy cluster strong lensing \citep{hennawi2007}. 
\citet{fedeli2006} conclude that accounting for mergers in simulations can nearly double 
the strong lensing optical depth for a subset of lenses -- those at z $\geq0.5$. However, 
a large fraction of cluster lenses lie at z $<0.5$, thus reducing the total 
magnitude of merger effects on simulated giant arc samples.

Having accounted for an exhaustive set of physical factors that could increase giant arc 
counts by making simulated clusters into better strong lenses, we are left with one 
hypothesis that has nothing at all to do with the cluster lens population. 
\citet{hamana1997} first argued that giant arc statistics depend sensitively on any
evolution of the galaxy luminosity function at high redshift, and \citet{oguri2003}
identify the uncertainty in the galaxy luminosity function at high redshift as a factor
that can significantly impact predictions for giant arcs based on their semi-analytic
models. This effect was quantified in simulations by placing background sources at
a series of different redshift planes and computing the integrated strong lensing
cross section for a simulated cosmological volume as a function of source redshift
\citep{dalal2004,wambs2004}. 

Interestingly, \citet{wambs2004} are able to
approximately reproduce the \citet{bartelmann1998} predictions by using a background
galaxy population placed entirely at z$_{s}=1$, but when background galaxies are placed at
z$_{s}=1.5$ they recover a factor of $\sim10$ increase in the number of predicted giant
arcs, and a factor of $\sim20$ increase for sources at z$_{s}=2$. \citet{dalal2004} also
identify a significant increase in the total strong lensing cross-section with increasing
source redshift, but measure the increased cross-section at higher source redshift to be
a factor of $\sim3$ smaller than \citet{wambs2004}. Simulations which evaluate lensing
efficiency with all sources placed at a single source plane are not physically realistic,
but the work by \citet{wambs2004} and \citet{dalal2004} emphasizes the importance of
understanding the distribution of source redshifts of background galaxies available to
be strongly lensed by foreground galaxy clusters. 

\subsection{These Results In the Context of Other Work}
\label{sec:context}

There is strong evidence that the total strong lensing cross-section for galaxy clusters 
can easily be increased by as much as a factor of at least $\sim7-10$ if a majority of 
the sources lensed into arcs reside at z $\gtrsim2$. Our results analyzing the optical 
colors for a large, well-selected sample of $105$ giant arcs provide a measurement 
of the distribution of redshifts for sources which are lensed into arcs, with 
$71^{+5}_{-4}~\%$ at z $\geq1.0$, $64^{+6}_{-4}~\%$ at z $\geq1.4$, $56^{+5}_{-4}~\%$ at 
z $\geq1.9$, and $21^{+4}_{-2}~\%$ at z $\geq2.7$. The inferred median redshift here is 
approximately $\bar{z}_{s} = 2.0$, which is consistent with the median redshift of $1.82$ 
determined from a much smaller sample in \citet{bayliss2011a}. Also encouragingly, 
the fraction of giant arcs measured at z $>[1.0,1.4,1.9]$ in this paper and 
in \citet{bayliss2011a} are consistent to within the statistical errors of the two 
measured redshift distributions. As a caveat to the sorting of arcs into precise 
redshift bins, we point out that the assumptions that go into 
defining regions in color space that associate with specific redshift ranges are 
likely to be much more robust in some cases than others. Specifically, the 
Lyman-$\alpha$ forest and the Lyman-limit break moving through the $u-$band 
at $z\gtrsim1.9$ will strongly affect the $u-g$ color of any galaxy regardless 
of the shape of its spectral energy distribution (SED). Color criteria at 
$z\lesssim1.9$, however, rely more heavily on the robustness of the assumptions 
made about the properties of the stellar populations within the observed galaxies.

With robust measurements of the giant arc redshift distribution for two samples of 
giant arcs, it is also important to confirm that the measured distributions are 
sensible in the context of other observations of high redshift galaxy populations. 
\citet{bayliss2011a} test the redshift distribution of $28$ 
bright giant arcs against simple physical models based on results from strong 
lensing simulations and galaxy catalogs from the COSMOS survey \citep{ilbert2009} 
and find good agreement with the observed redshift distribution. The form of the 
model redshift distribution is, 

$$\frac{dp_{\rm arc}}{dz_{s}} = \frac{\sigma_{\rm arc} \frac{dn}{dz_{s}}}{\int dz_{s} \sigma_{\rm arc} \frac{dn}{dz_{s}}} ~ ,$$

where the model uses a normalized strong lensing cross-section from \citet{fedeli2010} 
evaluated at a single source plane, z$_{s}=2$, and we solve for the giant arc 
cross-section at different source redshifts by assuming a universal density 
profile that is described by a power-law slope, $\alpha$, in the regions 
where strong lensing is observed. The number density of background galaxies as 
a function of redshift can be computed from the COSMOS catalogs for a given 
limiting magnitude and an estimate of the average magnification factor. We do 
not have good constraints on the average magnification of these sources, and 
the literature is not unanimous on the precise value of the slope of the 
density profile, $\alpha$, in the cores of clusters. However, by testing 
models that sample a range of reasonable values for these two parameters 
($-1.7 < \alpha < -1.3$, $26 < g_{lim} < 24$) we 
can produce synthetic redshift distributions that are in good agreement with 
the data \citep{bayliss2011a}. Though the broad bins that we define in this paper 
limit our ability to perform the same statistical comparisons that are used in 
\citet{bayliss2011a}, we can compare the fractions of arcs above various 
thresh-holds in the models and in our color-based redshift constraints and find 
good agreement. Because the sample analyzed in this paper was identified in imaging 
that is approximately $2$ magnitudes deeper than that used in \citet{bayliss2011a}, 
the fact that both redshift distributions can be well-described by the same simple 
physical models has an interesting implication; the primary difference between the 
brightest arcs selected from shallower imaging vs somewhat fainter arcs found in 
deeper imaging is that the brighter arcs are typically the cases with the most 
extreme magnifications.

\begin{figure}
\centering
\includegraphics[scale=0.55]{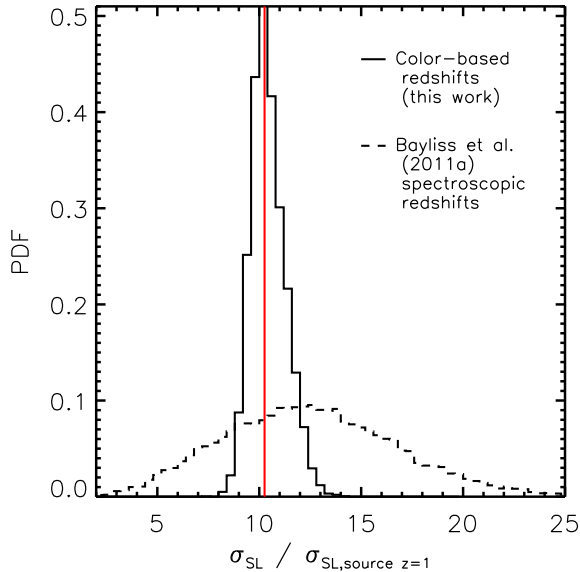}
\caption{\scriptsize{The probability distribution for the estimated increase in the
total cross-section for the production of gravitationally lensed arcs
with length-to-width ratios $\gtrsim5$, relative to the predictions
of \citet{bartelmann1998}. Estimates are made from the redshift
distribution derived in this paper using color-based redshift
assignments (solid line), as well as the redshift distribution
reported in \citet{bayliss2011a} (dashed line). The two estimates of
the increase in giant arc cross-section have nearly identical
expectation values of $\sigma_{SL}/\sigma_{SL,z=1} = 10.3$ --
indicated by the vertical red line -- but the much larger sample used
in this work vastly improve the precision of the estimate.}}
\label{sigma_sl}
\end{figure}

\subsection{Implications for Giant Arc Statistics}

With the giant arc redshift distribution now constrained in this paper, it is 
possible to calculate the approximate increase in the total cross-section for 
arc production compared to published predictions. We estimate the relative 
cross-section for different background source redshift distributions using 
Figure $1$ in \citet{wambs2004}, which provides a measurement of the strong 
lensing cross-section as a function of source redshift. The total cross-section 
for our measured redshift distribution is simply the integrated cross-sections 
from \citet{wambs2004} Figure $1$, weighted by the fraction of 
giant arcs determined to be within different redshift bins. The uncertainty in this 
determination is propagated from the color-based redshift identification, estimated by 
varying the weights in accordance with our measured uncertainties in the fraction of 
giant arcs located within each redshift bin (see Table~\ref{redshift_frac}).

The resulting total increase in the efficiency for giant arc production, normalized 
to the \citet{bartelmann1998} result with lensed sources at z$_{s}=1$, is 
$\sigma_{SL}/\sigma_{SL,z=1} = 10.3^{+1.1}_{-0.6}$. The uncertainty reported 
in this measurement does not incorporate an attempt to account for published evidence 
that the magnitude of the effect is smaller than \citet{wambs2004} claim 
\citep{dalal2004,li2005,fedeli2006}, but it is clear that the high-redshift nature 
of the lensed sources is a significant factor to be accounted for in predictions of 
giant arc abundances. We can compare our new estimate of $\sigma_{SL}/\sigma_{SL,z=1}$ 
to the same value estimated using the spectroscopic redshift distribution of 
\citet{bayliss2011a}, $\sigma_{SL}/\sigma_{SL,z=1} = 11^{+5}_{-3}$ (Figure~\ref{sigma_sl}). 
The improved constraints on the arc redshift distribution using a sample of $105$ -- 
along with the fact that four of the $28$ arcs analyzed in \citet{bayliss2011a} had 
only very broad redshift constraints spanning a range in $\Delta z = 1.8$ -- 
provides a much tighter estimate of the increase that we should expect in giant arc 
counts when using an empirically calibrated redshift distribution, relative to the 
simple z$_{s}=1$ used in \citet{bartelmann1998}. Modern efforts to simulate giant arc 
counts can incorporate the empirically determined giant arc redshift and brightness 
measurements that we provide in this work, and thus produce refined predictions for 
giant arc statistics. 

We have established that incorporating a background source population into future efforts 
to predict giant arc counts should increase the number of predicted arcs by as much as a 
factor of $\sim10$ beyond the results of \citet{bartelmann1998}, and thereby diffuse the 
tension between observations and prediction for arc counts. However, we have yet to 
discuss an important detail regarding most published work in which giant arcs are 
produced by ray-tracing simulations. Specifically, most simulations run in the past 
$\sim10-15$ years were run in cosmologies with a value for the normalization of the matter 
power spectrum, $\sigma_{8}$, that is higher -- typically $\sigma_{8}=0.9-0.95$ -- than the 
value that is preferred by current observational constraints of 
$\sigma_{8}\sim0.81\pm0.03$ \citep{komatsu2011,larson2011}. The precise value of 
$\sigma_{8}$ used in simulations is of particular importance for predictions of 
giant arc statistics because the high-end tail of the mass function varies 
sensitively with $\sigma_{8}$.

The quantitative impact of $\sigma_{8}\sim0.81$ on 
predictions of giant arc counts has been studied with N-body \citep{li2006} 
and semi-analytic techniques \citep{fedeli2008}, and both confirm the 
intuitive expectation that lower values of $\sigma_{8}$ result in significantly 
lower giant arc abundances. The difference between a cosmology with $\sigma_{8}=0.9$ 
vs $\sigma_{8}=0.74$ is found to be a factor of $\sim6$ in simulations \citep{li2006}, 
and semi-analytic modeling by \citet{fedeli2008} suggests that $\sigma_{8}\sim0.8$ 
under-predicts giant arc counts by an order of magnitude even after attempting to 
account for factors such as mergers and cluster substructure.

The \citet{fedeli2008} models do not account for contributions from galaxies embedded 
in cluster potentials, and use a background source description that is based on somewhat 
out-dated observations using the Hubble Deep Field \citep{casertano2000} that are subject 
to large cosmic variance uncertainties \citep{hennawi2007}, but the fact that an order 
of magnitude discrepancy persists in this work is clear motivation for the pursuit of 
modern simulations to better-characterize the expected giant arc abundance. The machinery 
is in place to produce these predictions in such a way that any persistent 
under-abundance of giant arcs in theory can be reasonably interpreted as evidence for 
real tension between the growth of structure in $\Lambda$CDM-- in the form of the abundance 
of the most massive clusters -- and the integrated strong lensing properties observed in 
the universe.  New, large giant arc samples are in the process of being identified from 
wide-field imaging survey data, and will soon be available in the literature 
(M. D. Gladders et al. 2011, in preparation; M. B. Bayliss et al. 2012, in preparation). 
These forthcoming samples will provide observational constraints that match the fidelity 
of the next generation of improved predictions, and set the stage for a definitive test 
of giant arc statistics.

\subsection{Other Applications for A Catalog of Lensed Galaxies At z $>2$}

The current literature contains a modest sample of extremely bright strongly lensed 
galaxies \citep[e.g.,][]{yee1996,smail2007,koester2010,wuyts2010}, some of 
which have been the target of detailed multi-wavelength follow-up studies 
\citep{fink2009,siana2009,quider2010}. Studies of individual galaxies at high
redshift are limited by the intrinsic faintness of the sources -- especially
where high fidelity observations of typical $\sim$L$^{*}$ galaxies are concerned. 
However, because faint galaxies (L$\lesssim$L$^{*}$) are far more
numerous than bright ones and therefore more likely to lie on the caustic
of foreground lensing clusters, we should expect most giant arcs to be 
intrinsically sub-L$^{*}$ galaxies.  It follows then that strongly lensed galaxies
are ideal targets for studying the properties of $typical$ high-z galaxies at
S/N that is unavailable in studies of galaxies in the field. By analyzing the optical 
broadband colors of $105$ arcs, we have shown that approximately half of all giant 
arcs formed around galaxy clusters can be expected to lie at z $\gtrsim2$. Forthcoming 
large catalogs of hundreds of galaxies that are strongly lensed by clusters should 
therefore provide of order hundreds of strongly lensed high-redshift sources that 
will be accessible for detailed individual study with the instrumentation available 
on $8-10$m class telescopes. 

\subsection{Summary}

We have measured $ugr$ magnitudes for a sample of $105$ giant arcs, and $z-$band 
magnitudes for $96/105$ of the same sample. The giant arc sample was identified 
in a systematic search of uniform depth $g-$band imaging data, and constitutes a 
representative subsample of two forthcoming large samples of giant arcs that 
are currently being identified in search of well-characterized cosmological 
search volumes. We then sort all $105$ arcs into five redshift bins based on 
well-established broadband color criteria, and thereby constraint the underlying 
redshift distribution of giant arcs identified with our selection method. The 
data indicate that z$_{s}=2.0\pm0.1$ is the typical redshift for these giant arcs, 
which agrees remarkably well with the redshift distribution measured from 
spectroscopic observations of a much smaller sample (N $=28$) giant arcs that 
were identified in much shallower imaging data. By establishing that approximately 
half of all bright giant arcs are galaxies at z $\gtrsim2$ we provide strong 
evidence for an easing of the tension between observations and predictions of 
the total abundance of giant arcs on the sky. 

\acknowledgments{
Support for this work comes in part from the Illinois Space Grant Consortium in the 
form of a graduate fellowship. I want to first give my sincere thanks the SOAR 
observing and support staff for their help in obtaining the observations that drove 
this paper. I also want to thank Michael Gladders, David Gilbank and Howard Yee for 
their work on RCS-2, Eva Wuyts for helpful discussions about giant arc photometry and 
SSP galaxy models, Felipe Barrientos and Mauricio Carrasco for graciously sharing 
unpublished spectroscopic redshifts, and Brad Holden for helpful discussion that 
expanded the scope of this work. I also want to thank my thesis committee members 
-- Stephen Kent, Andrey Kravstov and Richard Kron -- for their helpful comments. 
Lastly, I want thank Michael Gladders for his dedication and mentoring over the past 
four years.}

\bibliographystyle{apj}
\bibliography{bibtex_thesis}

\end{document}